\documentclass[10pt,aps,prb,twocolumn, nofootinbib]{revtex4-1}

\usepackage{natbib}
\usepackage{url}
\usepackage{amssymb,amsmath}
\usepackage{subfig}
\usepackage{graphicx}
\usepackage{color}
\usepackage{hyperref}
\usepackage{listings}
\hypersetup{
    bookmarks=true,         
    unicode=false,          
    pdftoolbar=true,        
    pdfmenubar=true,        
    pdffitwindow=false,     
    pdfstartview={FitH},    
    pdfauthor={Kevin Multani},     
    colorlinks=true,       
    linkcolor=black,          
    citecolor=black,        
}
\graphicspath{{figures/}}

\usepackage{siunitx}
\usepackage[utf8]{inputenc}
\usepackage{mwe}

\newcommand{\Abbref}[1]{Figure \ref{#1}}

\newcommand{\Tabref}[1]{Table \ref{#1})}
\begin{document}

\definecolor{dkgreen}{rgb}{0,0.6,0}
\definecolor{gray}{rgb}{0.5,0.5,0.5}
\definecolor{mauve}{rgb}{0.58,0,0.82}

\lstset{frame=tb,
  	language=Matlab,
  	aboveskip=3mm,
  	belowskip=3mm,
  	showstringspaces=false,
  	columns=flexible,
  	basicstyle={\small\ttfamily},
  	numbers=none,
  	numberstyle=\tiny\color{gray},
 	keywordstyle=\color{blue},
	commentstyle=\color{dkgreen},
  	stringstyle=\color{mauve},
  	breaklines=true,
  	breakatwhitespace=true
  	tabsize=3
}

\title{Influence of Te-doping on self-catalyzed VS InAs nanowires} 
\author{Nicholas A. Güsken\textsuperscript{1,2,3}, Torsten Rieger\textsuperscript{1,2}, Gregor Mussler\textsuperscript{1,2}, Mihail Ion Lepsa\textsuperscript{1,2}, Detlev Grützmacher\textsuperscript{1,2} \vspace{0.5cm}} 

\affiliation{\textsuperscript{1}\textit{Peter Grünberg Institute (PGI-9), Forschungszentrum Jülich, 52425 Jülich, Germany}\\
\textsuperscript{2}\textit{JARA-Fundamentals of Future Information Technology (JARA-FIT), Jülich-Aachen Research Alliance, Germany}\\
\textsuperscript{3}\textit{Present address: Department of Physics, Imperial College London, London, SW7 2AZ, U.K.}\\ }

\begin{abstract}We report on growth of Te-doped self-catalyzed InAs nanowires by molecular beam epitaxy on silicon (111) substrates. Changes in the wire morphology, i.e. a decrease in length and an increase in diameter have been observed with rising doping level. Crystal structure analysis based on transmission electron microscopy as well as X-ray diffraction reveals an enhancement of the zinc blende/(wurtzite+zinc blende) segment ratio if Te is provided during the growth process.
Furthermore, electrical two-point measurements show that increased Te-doping causes a gain in conductivity. Two comparable growth series, differing only in As-partial pressure by about $\SI{1e-5}{Torr}$ while keeping all other parameters constant, were analyzed for different Te-doping levels.
Their comparison suggests that the crystal structure is stronger affected and the conductivity gain is more distinct for wires grown at a comparably higher As-partial pressure.
\end{abstract}

\maketitle

\section{Introduction}

Nanowires (NWs) have attracted notably attention within the last decade as they are considered to constitute a promising building block for emerging and future technology. Their technical applications are diverse, ranging from field effect transistors and optical devices to solar cells \cite{Khanal2007,Breuer2011}. 
The broad applicability of NWs is based on their remarkable characteristics, such as a high aspect ratio, ultra low power dissipation and, in case of InAs, the absence of a Schottky barrier at the interface with metal contacts \cite{Xiang2006,Mohammad2011,Das2012}. The latter is due to the fact that InAs exhibits a surface accumulation layer, enabling ohmic contacting \cite{Piper2006}.\\
From the physics perspective, InAs NWs bear outstanding properties, i.e. a high electron mobility \cite{Dayeh2007a}, a low effective mass \cite{Dayeh2010}, a large g-factor \cite{Bjork2005} and strong Rashba spin-orbit coupling \cite{Liang2012,EstevezHernandez2010}. Due to this, they became an important ingredient regarding quantum information related research \cite{Das2012,Nadj-Perge2010,Leijnse2013,Lutchyn2010}.\\
InAs NWs are commonly grown via gold droplet catalyzed growth. However, the use of Au-catalysts is not compatible with silicon based technology \cite{Breuer2011,Mohammad2011,Du013,Allen08}. Thus, catalyst-free growth in the vapor-solid (VS) mode was applied within this communication \cite{Rieger2013}.\\
In contrast to Au-droplet-assisted growth, the VS mode results in a high occurrence of stacking faults within the NW. This means a frequent switching between the zinc blende (ZB) and the wurtzite (WZ) structure within the growth direction of the wire \cite{Hertenberger2011,Wei2009,Dimakis2011}. The change in crystal stacking is called polytypism and constitutes a major drawback of the self-catalyzed growth as it degrades the transport \cite{Thelander2011,Schroer2010} and alters the optical properties \cite{Heiss2011}.
One way to counteract the diminished charge transport is the use of doping, i.e the incorporation of additional carriers. \\
However, the well-established doping methods used for semiconducting layers can not simply be transfered to the wire structures \cite{Erwin2005,Ghoneim2012}. For instance, unintentional surface doping leading to non-uniform carrier incorporation along the wire can occur \cite{Allen2009}. This type of doping describes a predominant but inhomogenous incorporation of dopants close to the wire surface, leading to a resistivity gradient along the axial and radial direction. The common dopants Si and C exhibit an amphoteric behavior. They can act as n- or p-type dopant, depending on how they are incorporated into the material \cite{Piper2006,Ghoneim2012,Thelander2010}, impeding precise control. \\
Tellurium-doping (Te-doping) in VS grown GaAs NWs has been reported to have an impact on the wire morphology and crystal structure \cite{Suomalainen2015}.
This publication investigates Te-doping in InAs NWs, thus extending the NW doping toolbox. 
Moreover, this communication provides information regarding the impact of doping on the NW morphology and the switching between the ZB and the WZ structure within the VS growth in the presence of Te. Investigations based on scanning electron microscopy (SEM) disclosed a strong impact of Te on the NW morphology. High resolution transmission electron microscopy (HR-TEM) \cite{ErnstRuska-Centre2016} and X-ray diffraction (XRD) measurements served to evidence a change in the ZB/(WZ+ZB) ratio and electrical two point measurements showed an increase in conductivity with raising Te-doping level.

\section{Experimental}

InAs NWs were grown in the VS mode without the use of any foreign catalyst on n-type Si(111) substrates.
Prior to growth, the substrates were cleaned using HF and DI-water. A consecutive hydrogen peroxide treatment for $\SI{45}{seconds}$ leads to the formation of a few angstrom thick $\mathrm{SiO_2}$ film containing pinholes, which serve as nucleation centers for the NW growth \cite{Rieger2013}. After the oxidation, the substrates were immediately transferred into the load-lock in which they were heated to $\SI{200}{\degree C}$ for $\SI{45}{minutes}$. This was followed by a degassing step within the preparation chamber, heating the samples at $\SI{400}{\degree C}$ for another $\SI{45}{minutes}$.\\
The NWs were grown at a substrate temperature of $\SI{475}{\degree C}$ for 1:20$\hspace{0.05cm}\mathrm{h}$ in an Omicron Pro 100 molecular beam epitaxy (MBE) chamber.
An In-growth rate of $\SI{0.1}{\micro m h^{-1}}$ was used for the NW growth. Arsenic was provided via an As cracker-cell
and the $\mathrm{As}_4$-beam equivalent pressure (BEP) was adjusted to values of $\SI{2.3e-5}{Torr}$ and $\SI{3.3e-5}{Torr}$. The first sample series (series A) was grown at higher As-partial pressure compared to a second growth series (series B) (cf. \Tabref{GParam}), while keeping all other parameters constant.
Tellurium was supplied during the growth using stoichiometric GaTe.
The operation temperature of the effusion cell was varied between $\SI{401}{\degree C}$ and $\SI{562}{\degree C}$ based on calibrations conducted on Te-doped GaAs layers via Hall-measurements. The GaTe-cell temperatures $\SI{401}{\degree C}$, $\SI{447}{\degree C}$, $\SI{500}{\degree C}$ and $\SI{561}{\degree C}$ correspond to a carrier concentration of about $\SI{1e15}{cm^{-3}}$, $\SI{4e16}{cm^{-3}}$, $\SI{5e17}{cm^{-3}}$ and $\SI{6e19}{cm^{-3}}$ respectively, in GaAs (100) layers used for calibration.
In order to process two point contacts, the NWs were mechanically transferred on a pre-patterned Si substrate which was covered with $\SI{200}{nm}$ $\mathrm{SiO_2}$. Prior to metal deposition the wires were spin-coated by a three layer system of 50K (AR-P639.04), 50K, and 950K (AR-P679.04) PMMA resist upon which the contact shape was defined via e-beam lithography. After development the contact area was passivated by diluted $3.5\%$ ammonium polysulfide ($\mathrm{H_2O:(NH_4)_2S_3}$, 34:1) at $\SI{60}{\degree C}$ for $\SI{30}{minutes}$. The electrodes, consisting of $\SI{100}{nm}$ titanium and $\SI{40}{nm}$ gold were deposited via an electron beam evaporator.
The complete list of samples investigated via SEM, TEM, XRD and electrical measurements is presented in \Tabref{GParam}. Here, the letters A,B,C indicate the sample series which were grown each at different As-partial pressures but apart from that under equal conditions. A GaTe temperature of $\SI{0}{\degree C}$ corresponds to a closed cell shutter.

\captionof{table}{$\mathrm{As}_4$-BEP and GaTe-cell doping temperature of the samples used for analysis.}
 \begin{center}
    \begin{tabular}{ l|c c}
Sample &{ $\mathrm{As}_4$-BEP} \hspace{-2cm}&{ GaTe Temp.}\\&[$\SI{e-5}{Torr}$] &[\SI{}{\degree C}]\\ \hline
A1& 3.3 &0\\ 
B1& 2.3 &0\\
A2& 3.3 &401\\
B2& 2.3&401\\
A3& 3.3 &447 \\
B3&2.3 &447 \\
B4&2.3&462 \\
A4&3.3&500 \\
C1&3&500 \\
A5& 3.3&561 \\
\end{tabular}\label{GParam}
\end{center}

\section{Results}

\subsection{Morphology}

\begin{figure}[]
\begin{center}
\includegraphics[scale=0.56]{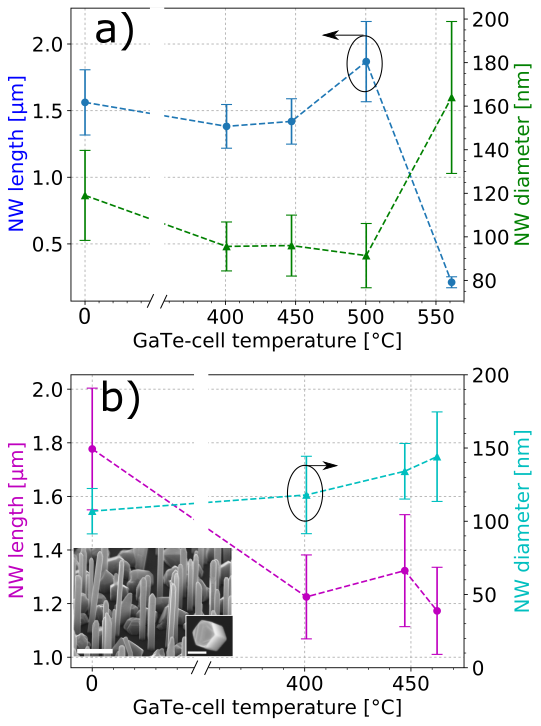}
\caption{Nanowire mean length and diameter at different GaTe-cell temperatures. a) Series A was grown at an As-partial-pressure of about $\SI{3.3e-5}{Torr}$. b) Series B was grown at an As-partial-pressure of $\SI{2.3e-5}{Torr}$. The broken lines are guidance for the eyes. The SEM micrograph shown in the inset depicts undoped InAs NWs surrounded by crystallites formed during growth. The scale bar is $\SI{300}{\nm}$ and $\SI{120}{\nm}$, respectively.}
\label{Morphology}
\end{center}
\end{figure}

SEM imaging was used in order to investigate the influence of Te-doping on the wire morphology. The results are presented in \Abbref{Morphology}. Every data point on the graphs constitutes the mean of at least 40 wires and the error bar their standard deviation.

\Abbref{Morphology} {a)} shows the morphology of wire series A grown at an As-partial-pressure of about $\SI{3.3e-5}{Torr}$. The GaTe-cell temperature ranged from $\SI{0}{\degree C}$ to $\SI{561}{\degree C}$. Taking the error bar into account, no distinct trend of the NW diameter and length is observed until a cell temperature of $\SI{500}{\degree C}$. However, at $\SI{561}{\degree C}$ the supply of Te is clearly detrimental, leading to a strong increase in diameter and decrease in NW length.
Growth series B, depicted in \Abbref{Morphology} {b)} has been grown at a comparably lower As-pressure of $\SI{2.3e-5}{Torr}$. The inset shows exemplary a SEM sideview of a grown sample, exhibiting InAs NWs and clusters on the substrate surface.  Here, a GaTe-cell temperature range from $\SI{0}{\degree C}$ to $\SI{462}{\degree C}$ was explored. We observe a decrease in length when Te is added during growth for series B. Comparing the measurements of series A and B in the same temperature interval, it is observed that in particular the decrease of NW length is more distinct at comparably lower As-pressures (series B). However, the same overall trend, i.e. a decrease in NW length and increase in diameter is observed for both series.\\
Si doping leads similary to an increased diameter and decreased length for InAs and GaAs, independently of the growth process (MBE or metalorganic vapour phase epitaxy (MOVPE)) \cite{Dimakis2013,Wirths2011,Ghoneim2012}.
The same change in dimensions was observed for Te-doping of self-catalyzed GaAs NWs grown by MBE \cite{Suomalainen2015}. It seems that independent of the used material system, i.e. III-V materials doped with group IV (InAs/Si, GaAs/Si) or group VI materials (GaAs/Te, InAs/Te), the same overall trend regarding morphology is observed. \\
Te exhibits a rather large covalent radii with respect to the host lattice atoms and acts therefore as surfactant \cite{Wixom2004,Caroff2011}. Thus we attribute the observed behavior to a decrease in diffusivity of the In atoms caused by the surfactant Te. This in turn causes the increase in radial growth and a decrease in length as the In adatoms are hindered on their way to the NW tip where they control the growth \cite{Dimakis2013}. However, comparing \Abbref{Morphology} {a)} (series A) and \Abbref{Morphology} {b)} (series B) we find that the As-pressure influences how the wire morphology is affected by the Te addition.
The finding suggests that it might be possible to counteract the diminishing impact of Te on the radial and axial dimension of the InAs wire by increasing the As pressure to a certain extent.

\subsection{Crystal structure}

\begin{figure}[h!]
\begin{center}
  \includegraphics[scale=0.48]{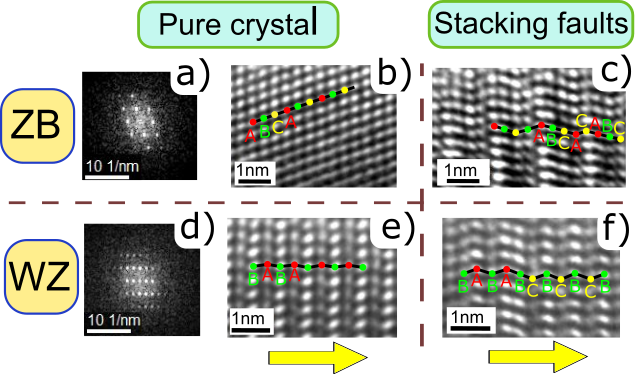}
  \caption{HR-TEM images of InAs NWs, illustrating the ZB and WZ crystal structures with and without stacking faults. The yellow arrows indicate the [111] growth direction. The colored dots and the black lines are guidance for the eyes to retrace the stacking characteristic. a) FFT diffraction pattern for defect-free ZB. b), c) ZB structure. d) FFT diffraction pattern for defect-free WZ and e)-f) WZ structure.}
  \label{TEM}
\end{center}
\end{figure}
The impact of the Te-dopants on the crystal structure was investigated using TEM and XRD.
Adopting the classification used by Caroff et al.\cite{Caroff2011}, a crystal stacking sequence was assigned to a ZB (cf. \Abbref{TEM} {a)})  or a WZ (cf. \Abbref{TEM} {d)}) segment if the stacking sequence followed exactly four bilayers of atoms. This means ...ABCA... was counted as a ZB segment and ...ABAB... as a WZ segment. This is illustrated in \Abbref{TEM} {b)} and \Abbref{TEM} {e)}. Here, every letter represents a bilayer of atoms. 
Some wire sections are interrupted by stacking faults (SFs) consisting of a missing or excess layer within the crystal sequence, as presented in \Abbref{TEM} {c), f)}. Albeit rarely observed, rotational twinning is present in some segments as well (not shown here).

\begin{figure*}[]
\begin{center}
\includegraphics[scale=0.30]{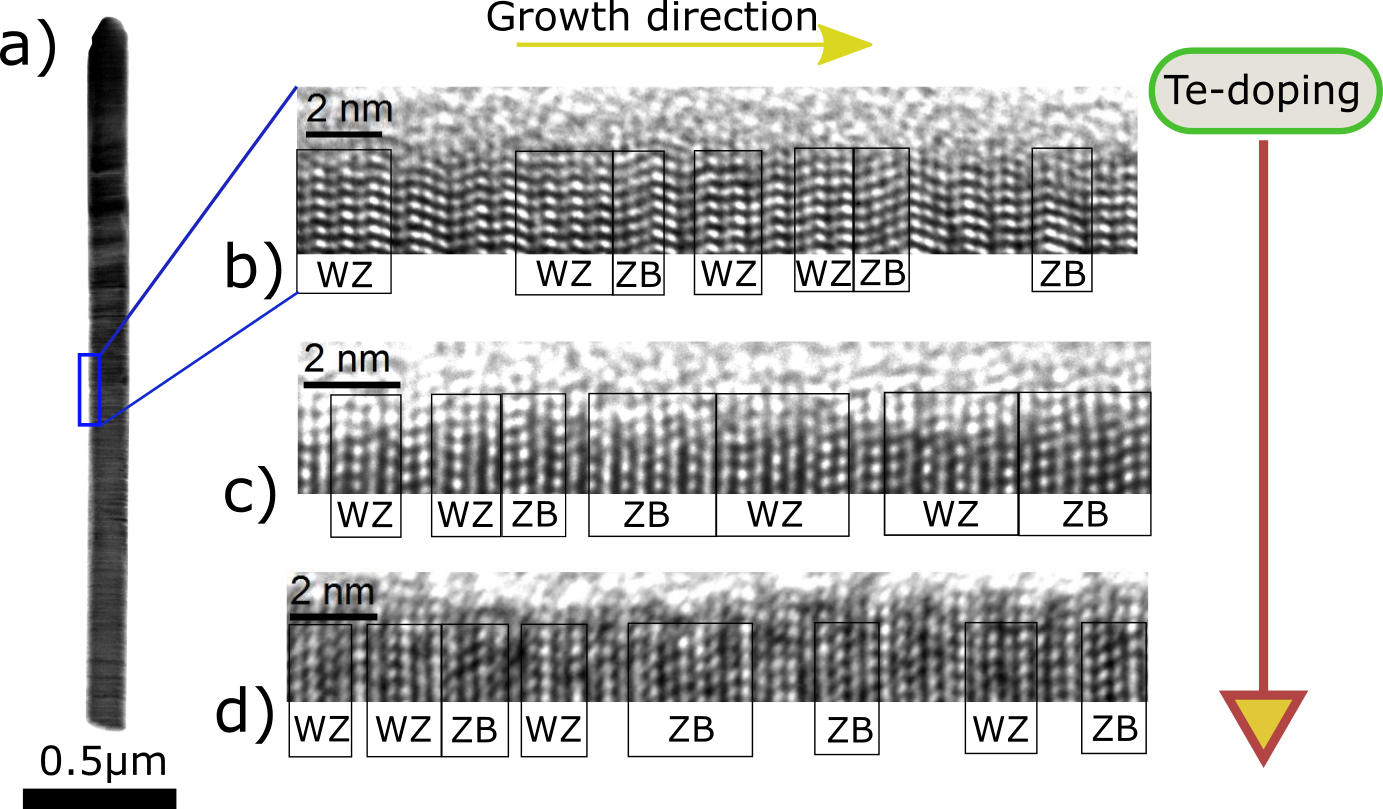}
\caption{TEM images depicting the crystal structure in undoped and Te-doped InAs NWs. a) Side view of an InAs NW. b)-d) HR-TEM images of the InAs NW crystal structure (image turned $\SI{90}{\degree}$ clockwise). The WZ and ZB areas are labeled. The following samples and GaTe-cell temperatures were chosen: b) B1, undoped, i.e. $\SI{0}{\degree C}$. c) B3, $\SI{447}{\degree C}$. d) C1, $\SI{500}{\degree C}$.}
\label{TEMCrystalDopingInfluence}
\end{center}
\end{figure*}

\begin{figure}[h!]
\begin{center}
\includegraphics[width=8.3cm]{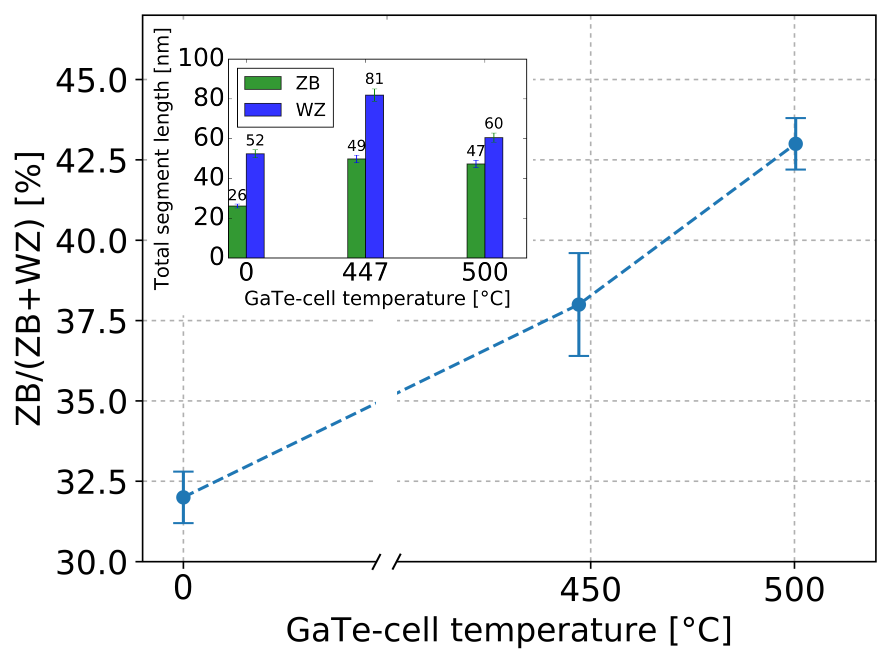} 
\caption{Ratio of the number of ZB segments and the total number of segments identified as WZ or ZB in dependence on the GaTe-cell temperature. For the first two measurements B1 and B3 have been analyzed. At $\SI{500}{\degree C}$ the results of wires C1 and A4 were merged as they were grown at similar As-BEP. The bar plot in the inset depicts the accumulated length of all WZ and ZB segments present in the NW at the indicated cell temperature, respectively. About \SI{150}{nm} of the NW were analyzed at each temperature according to the classification explained in the text.}
\label{TEMAnalysisCurve}
\end{center}
\end{figure}

Crystal sections were identified as ZB or WZ segments only if one full sequence consisting of four bilayers of atoms was observed. The remaining sections were attributed to SFs or rotational twins.\\ The crystal structure at three different doping levels evaluated according to the explained characteristics is illustrated in \Abbref{TEMCrystalDopingInfluence}. Different WZ and ZB areas are highlighted. However, for the analysis only individual segments were counted.
In order to quantify the influence of the Te-doping onto the NW crystal structure, a total segment of about $\SI{150}{nm}$ for each doping level was analyzed (cf. \Abbref{TEMCrystalDopingInfluence} b)-d)). The ZB/(WZ+ZB) segment ratio was determined by counting the number of single ZB and WZ segments. The samples B1, B3, C1 and A4 at $\SI{0}{\degree C}$, $\SI{447}{\degree C}$ and $\SI{500}{\degree C}$ were analyzed (cf. \Abbref{TEMAnalysisCurve}), respectively. We observe an enhancement of the ZB/(WZ+ZB) segment ratio with increasing GaTe-cell temperature. This trend is illustrated in \Abbref{TEMAnalysisCurve}. Comparing the first two data points ($\SI{0}{\degree C}$ and $\SI{447}{\degree C}$), the enhanced ratio is due to a stronger increase in ZB segments in comparison to the increase of WZ segments from the undoped to the lowest doping temperature (cf. inset \Abbref{TEMAnalysisCurve}). Both structure types are enhanced and the number of SFs is decreased. However, the trend differs for the third point. When comparing the highest with the lowest doping level ($\SI{500}{\degree C}$ and $\SI{447}{\degree C}$), we find that the number of WZ segments decreases and the number of ZB segments stays almost constant (cf. inset \Abbref{TEMAnalysisCurve}) while the number of SFs increases. This leads to a raised ratio. Still, the ZB section is promoted in comparison to the undoped case. Finally, the findings show that the Te-doping indeed enhances the ZB/(WZ+ZB) segment ratio. However, it remains ambiguous if the formation of ZB segments is strictly promoted by Te incorporation.

In order to complement the observations made by TEM, XRD measurements were performed.

\begin{figure}[]
\center
\includegraphics[scale=0.11]{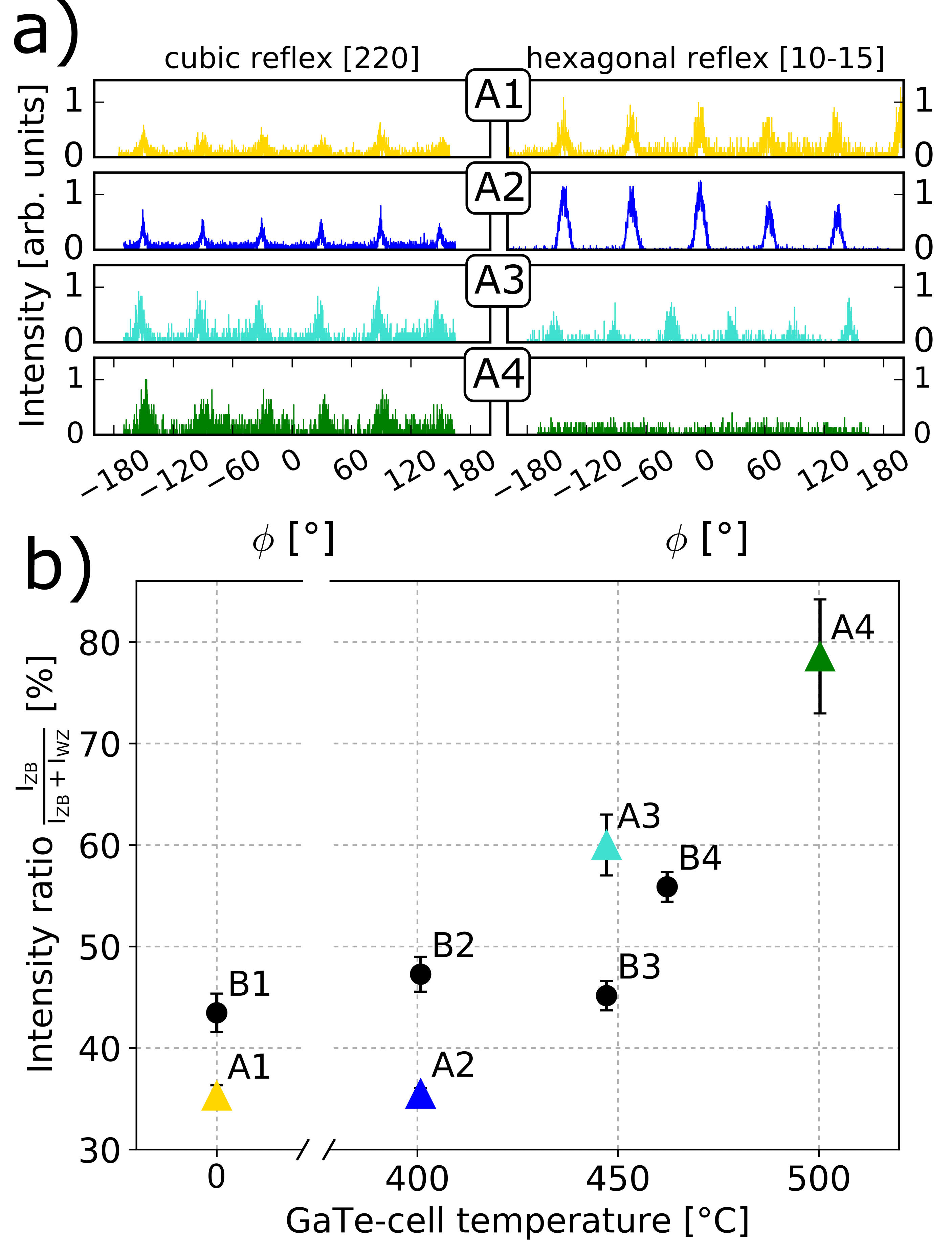}
\caption{a) $\phi$ scans obtained via X-Ray measurements on InAs NWs A1-A4. b) Resulting $\mathrm{I_{ZB}/(I_{WZ}+I_{ZB})}$ intensity ratio vs. GaTe-cell temperature. The triangular data points are extracted from the measurements depicted in a) for sample A1-A4. The black dots indicate the data points of samples B1-B4.}
\label{GregorResults}
\end{figure}

We conducted $\phi$-scans focusing on the cubic [220] and the hexagonal $[10\bar{1}5]$ reflexes. These reflexes can be unambiguously attributed to the ZB and the WZ structure, respectively. The measurement of the respective intensities allowed to extract the $\mathrm{{I_{ZB}}/({I_{ZB}+I_{WZ}})}$ intensity ratio. The $\phi$-scans depicted in \Abbref{GregorResults} { a)}, served to determine the relative intensities of the ZB and the WZ peaks at each GaTe-cell doping temperature. For the ZB reflex, six peaks occur even though the cubic lattice should only lead to a 3 fold symmetry. We assign these peaks to symmetric twins in the ZB structure. The six fold symmetric peaks occurring in the WZ scan are characteristic for the hexagonal crystal structure and match our expectations. Here, the signal intensity of InAs surface crystallites (cf. inset in \Abbref{Morphology}) is assumed to be two orders of magnitude smaller than the NW signal \cite{Sourribes2014a} and can be thus neglected. 
The corresponding intensity ratio $\mathrm{{I_{ZB}}/({I_{ZB}+I_{WZ}})}$ is plotted in \Abbref{GregorResults}{ b)} (colored triangles). It shows an increase of the $\mathrm{{I_{ZB}}/({I_{ZB}+I_{WZ}})}$ intensity ratio with increasing GaTe-cell temperature. 
This result is in accordance with the observation already obtained from the TEM analysis.
Note that the given intensity ratios do not represent the real ZB/WZ proportion but constitute a qualitative result. This is due to the fact that different reflexes are of different intensity, according to the structure factor which has not been taken into account explicitly. However, the comparison among the data points remains valid.\\
The same reflex-sensitive measurement was conducted for series B which was grown at a lower As pressure than series A presented above. The results depicted in \Abbref{GregorResults}{ b)} (black dots) show a similar trend as the A series, i.e. an increase in the ZB/(WZ+ZB) intensity ratio. However, the impact of the Te atoms on the crystal structure is less distinct in comparison to higher As pressures. This observation suggests that the As atoms facilitate the incorporation of Te atoms, which in turn leads to a change in crystal structure. Hence, a stronger impact on the ZB/(WZ+ZB) ratio is observed for the respective higher As pressure. The decrease of the intensity ratio at $\SI{447}{\degree C}$ in \Abbref{GregorResults}{ b)} (black dots) might be due to shadowing effects as the NW density for sample B3 was above average, though this is not yet fully understood.\\
From the TEM results presented above, one concludes that the NWs which were grown under Te supply show an increased number of ZB and WZ segments and hence less SFs compared to the undoped case. Further, the XRD measurements indicate that the ZB/(WZ+ZB) intensity ratio increases with increasing Te-doping level which is qualitatively in line with the TEM measurements made. In contrast to common elements used for doping of III-V materials such as Si (InAs/Si \cite{Dimakis2013}, GaAs/Si), C (GaAs/C) or Be (GaAs/Be), Te clearly affects the crystal structure of the NW. The promotion of the observed ZB formation might originate from a change of surface energies, lowering the energy barrier for ZB nucleation. This was equally observed in zinc doped InP nanowires \cite{Algra2008} where Au-catalyzed VLS growth was used. However, further research is needed to clarify the underlying mechanism.

\subsection{Electrical measurements} 

\begin{figure}[h!]
\center
\includegraphics[scale=0.55]{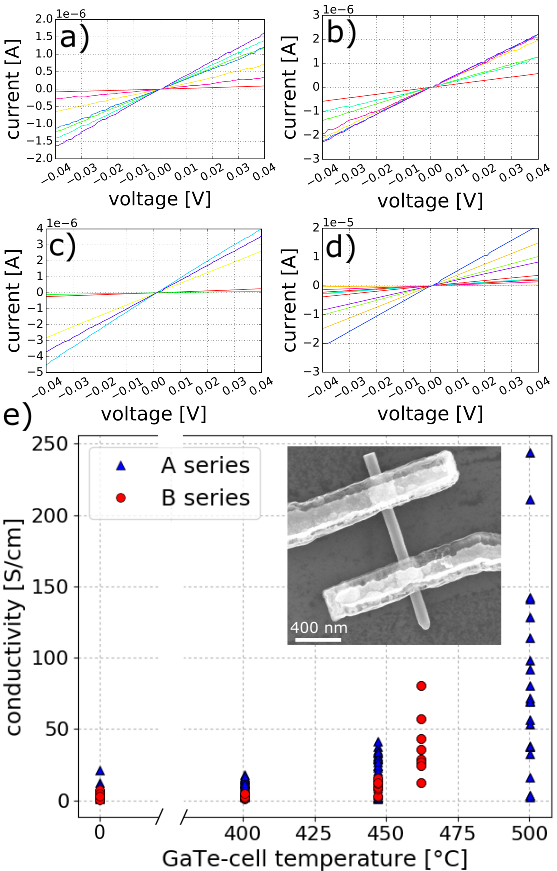} 
\caption{a-d) Exemplary I-V measurements of InAs NWs at a GaTe-cell temperature of $\SI{0}{\degree C}$, $\SI{401}{\degree C}$, $\SI{447}{\degree C}$ and $\SI{500}{\degree C}$ (series A) measured via two point contacts. e) Determined conductivity values of Te-doped InAs NWs in dependence of GaTe-cell temperature for the A (high As-BEP) and B (low As-BEP) series. The inset shows Ti/Au two point contacts on top of an InAs NW used for the transport measurements.}
\label{conductivities}
\end{figure}

The conductivity defined by $\sigma=\mathrm{A} \cdot \mathrm{R} \cdot \mathrm{L_w^{-1}}$ was extracted from two point measurements using Ti/Au contacts. Here, A is the hexagonal cross section of the wire with $\mathrm{A}=3\sqrt{3}\mathrm{d^2_{NW}}/8$ where $\mathrm{d_{NW}}$ is the maximal diameter, R the resistance and $\mathrm{L_w}$ the distance between the electric contacts. $\mathrm{L_w}$ and $\mathrm{d_{NW}}$ have been measured individually for every wire via SEM imaging. 
Exemplary I-V characteristics of undoped and doped InAs NWs are presented in \Abbref{conductivities} {a)-d)}. The graphs show the expected ohmic behaviour due to the surface accumulation layer of InAs \cite{Piper2006, Noguchi1991}. \\The conductivity in dependence on the doping level was determined based on the I-V measurements and the NW geometry. 
The resulting dependence between the conductivity and the GaTe-cell temperature is illustrated in \Abbref{conductivities} {e)}. At each temperature at least 20 NWs were examined for series A (blue dots). Comparing undoped and doped wires, an increase in conductivity of about one order of magnitude at the highest doping level is observed. At a GaTe-cell temperature of $\SI{500}{\degree C}$ an average conductivity of about $\SI{80}{S/cm}$ was determined (compared to about $\SI{8}{S/cm}$ for $\SI{0}{\degree C}$). This result shows that incorporation of Te indeed has a strong impact, enhancing the conductivity. \\
The comparison of the XRD and the conductivity measurement suggests that below $\SI{401}{\degree C}$ the impact of Te on the crystal structure and transport properties plays merely a minor role.

Our measurements show a large variance in conductivity at each doping temperature.
The large spread in conductivity of InAs NWs has been similarly reported in literature \cite{Potts2015}. No trend of the conductivity was observed with respect to a change in NW diameter or contact spacing, as expected \cite{Thelander2011}. Thus, we exclude the differing aspect ratio as source of error. We identify three main reasons responsible for the strong variance in conductivity: i) The contact passivation method using ammonium polysulfide could lead to an heterogeneous contact quality. ii) The wire surface is not passivated and surface states can be influenced by an inhomogeneous saturation of the dangling bonds at the wire side facets via water and oxygen which finally results in nonuniform surface oxidation. This in turn has a strong impact on the transport characteristics, leading to large errors \cite{Dayeh2007}. One way to prevent these heterogenous surface states is the passivation via in situ deposition of $\mathrm{Al_2O_3}$ \cite{Campabadal2011, Potts2015}.
iii) Inhomogeneous doping along the NW, as observed for Si doping \cite{Allen2009}, might also cause the large data spread, although we tried to exclude that by placing the contacts centered for each wire. Finally, variations in NW length (cf. \Abbref{Morphology}) and density can lead to shadowing effects, preventing from homogeneous Te incorporation across the sample. However, more systematic investigations are needed to identify the exact source of the large variance observed.\\
Additionally, conductivity measurements for NWs of series B grown at a comparably lower As pressure were conducted. Here, at least six wires were measured for each GaTe-cell temperature. The results depicted in \Abbref{conductivities} {e)} (red triangles) show a similar behavior as the ones discussed above for series A (blue dots). The conductivity of InAs NWs is increased for higher GaTe-cell temperatures. However, the effect is less distinct in comparison with series A, grown at a higher As pressure. Comparing the conductivities of both series at $\SI{401}{\degree C}$ and $\SI{447}{\degree C}$ in \Abbref{conductivities} {e)}, we find that the values for series A are about twice as large as the ones found for B. The XRD results presented above (cf. \Abbref{GregorResults}) illustrate that the crystal structure in series A is more strongly affected by the Te incorporation than in series B. The combination of both findings indicates that the raised conductivity is related to the change in crystal structure, i.e. the increased ZB/(WZ+ZB) intensity ratio. \\
It is known from literature that a modification in the InAs NW crystal structure from WZ dominated towards ZB dominated, enhances  conductivity \cite{Sourribes2014a,Potts2015,Fu2016}. Based on TEM investigations on $\mathrm{InAs_{1-x}Sb_{x}}$ NWs Sourribes et al. reported an increase in conductivity by 1.5 for a gain in NW ZB fraction from $20\%$ to $80\%$ \cite{Sourribes2014a}. Our  TEM results (cf. \Abbref{TEMAnalysisCurve}) show a raised ZB/(WZ+ZB) ratio from $32\%$ (undoped NW) to $43\%$ (maxmimum doped NW) while the averaged conductivity value increases by about a factor of 10. This comparison suggests that the altered crystal structure is not the only reason for the conductivity enhancement. Although the modification of the crystal structure affects the carrier transport, the effect observed is probably likewise due to an augmented carrier density induced by Te acting as a donor.

\section{Conclusion}
In summary we have grown Te-doped self-catalyzed InAs NWs on Si (111) substrates via the vapor solid growth method. Te was provided by a GaTe-cell enabling the growth of Te-doped InAs NWs at different doping levels by adjusting the cell temperature. Two sample series grown at different As-BEPs were characterized by SEM, TEM, XRD and electrical measurements.\\
We have shown that Te changes the NW morphology leading to an overall trend of an increased radial and decreased axial growth rate. The impact is stronger at comparably lower As-partial pressures. We attribute the observed growth behavior to the surfactant behavior of Te impeding the axial growth controlled by In.\\ TEM and XRD measurements disclosed that the NW crystal structure is affected by Te addition, resulting in an increase of the ZB/(WZ+ZB) ratio for both growth series. The influence on the NW crystal structure grown at comparably higher As-BEP was more enhanced than observed for NWs grown at lower As-BEP. \\
Electrical two point measurements demonstrated an increase in conductivity for wires grown under Te supply. This was observed for two growth series, grown at different As pressures. The comparison between the two sample series showed that the crystal and electrical properties of InAs NWs are more strongly affected by Te addition at higher As pressures. The result indicates that the improved conductivity is strongly related to the change in crystal structure, i.e. the increase in ZB/(WZ+ZB) ratio. We attribute the enhanced transport properties to both the incorporated group VI element Te acting as a donor as well as an altered crystal structure. \\
This work constitutes an important contribution in order to extend the options for NW doping which is of great interest to counteract the degradation of transport properties by SFs.\\

{\bfseries Acknowledgement}: The authors want to thank Christoph Krause as well as Benjamin Bennemann for the support at the MBE cluster. Furthermore we thank the entire staff of the Helmholtz Nano Facility providing crucial assistance for the sample processing. In particular we acknowledge the work of Stefan Trellenkamp providing electron beam lithography.

\bibliography{library}
\end{document}